\documentclass[UKenglish,twocolumn,amsmath,amssymb, prb, aps, floatfix,superscriptaddress, 10pt]{revtex4-2}
\usepackage{float}

\usepackage{multirow}
\usepackage{amsmath,amssymb}
\usepackage[colorlinks=true,bookmarks=false,citecolor=blue,urlcolor=blue]{hyperref} %pdflatex
\usepackage{listings}
\usepackage{color}
\usepackage{graphicx}
\usepackage{caption}
\usepackage{physics}
\usepackage{braket}
\usepackage{ulem}
\usepackage{verbatim}
\usepackage{hyperref}
\hypersetup{colorlinks}
\usepackage{fancyhdr}

\usepackage{orcidlink} % ORCID icon package (RevTeX supports it)
\usepackage{xcolor}
\usepackage{tabularray}

\begin{document}

\title{Monitoring the generation of photonic linear cluster states
with partial measurements}

\affiliation{Centre for Nanosciences and Nanotechnologies CNRS, Université Paris-Saclay, UMR 9001, 91120 Palaiseau, France}
\affiliation{Quandela SAS, 10 Boulevard Thomas Gobert, 91120 Palaiseau, France}
\affiliation{Racah Institute of Physics, Hebrew University of Jerusalem, Jerusalem 91904, Israel}
\affiliation{These authors contributed equally to this work.}

\author{Valentin Guichard}
\email{valentin.guichard@quandela.com}
\affiliation{Centre for Nanosciences and Nanotechnologies CNRS, Université Paris-Saclay, UMR 9001, 91120 Palaiseau, France}
\affiliation{Quandela SAS, 10 Boulevard Thomas Gobert, 91120 Palaiseau, France}
\affiliation{These authors contributed equally to this work.}
\author{Leonid Vidro }
\affiliation{Racah Institute of Physics, Hebrew University of Jerusalem, Jerusalem 91904, Israel}
\affiliation{These authors contributed equally to this work.}
\author{Dario A. Fioretto\,\orcidlink{0000-0003-3829-4000}}
\affiliation{Centre for Nanosciences and Nanotechnologies CNRS, Université Paris-Saclay,
UMR 9001, 91120 Palaiseau, France}
\affiliation{Quandela SAS, 10 Boulevard Thomas Gobert, 91120 Palaiseau, France}
\author{Petr Steindl\,\orcidlink{0000-0001-9059-9202}}
\affiliation{Centre for Nanosciences and Nanotechnologies CNRS, Université Paris-Saclay, UMR 9001, 91120 Palaiseau, France}
\author{Daniel Istrati}
\affiliation{Racah Institute of Physics, Hebrew University of Jerusalem, Jerusalem 91904, Israel}
\author{Yehuda Pilnyak}
\affiliation{Racah Institute of Physics, Hebrew University of Jerusalem, Jerusalem 91904, Israel}
\author{Mathias Pont}
\affiliation{Centre for Nanosciences and Nanotechnologies CNRS, Université Paris-Saclay,
UMR 9001, 91120 Palaiseau, France}
\affiliation{Quandela SAS, 10 Boulevard Thomas Gobert, 91120 Palaiseau, France}
\author{Martina Morassi}
\affiliation{Centre for Nanosciences and Nanotechnologies CNRS, Université Paris-Saclay, UMR 9001, 91120 Palaiseau, France}
\author{Aristide Lemaître}
\affiliation{Centre for Nanosciences and Nanotechnologies CNRS, Université Paris-Saclay, UMR 9001, 91120 Palaiseau, France}
\author{Isabelle Sagnes}
\affiliation{Centre for Nanosciences and Nanotechnologies CNRS, Université Paris-Saclay, UMR 9001, 91120 Palaiseau, France}
\author{Niccolo Somaschi}
\affiliation{Centre for Nanosciences and Nanotechnologies CNRS, Université Paris-Saclay, UMR 9001, 91120 Palaiseau, France}
\author{Nadia Belabas}
\affiliation{Centre for Nanosciences and Nanotechnologies CNRS, Université Paris-Saclay, UMR 9001, 91120 Palaiseau, France}
\author{Hagai Eisenberg}
\affiliation{Centre for Nanosciences and Nanotechnologies CNRS, Université Paris-Saclay, UMR 9001, 91120 Palaiseau, France}
\author{Pascale Senellart\,\orcidlink{0000-0002-8727-1086}}
\affiliation{Centre for Nanosciences and Nanotechnologies CNRS, Université Paris-Saclay, UMR 9001, 91120 Palaiseau, France}

\begin{abstract}
Quantum states of light with many entangled photons are key resources for photonic quantum computing and quantum communication. In this work, we exploit a highly resource-efficient generation scheme based on a linear optical circuit embedding a fibered delay loop acting as a quantum memory. The single photons are generated with a bright single-photon source based on a semiconductor quantum dot, allowing to perform the entangling scheme up to 6 photons. We demonstrate $2$, $3$, $4$ and $6$-photon entanglement generation at respective rates of $6$kHz, $120$Hz, $2.2$Hz, and $2$mHz, corresponding to an average scaling ratio of $46$.  We introduce a method for real-time control of entanglement generation based on partially post-selected measurements. The visibility of such measurements carries faithful information to monitor the entanglement process, an important feature for the practical implementation of photonic measurement-based quantum computation.\end{abstract}
\maketitle

\section*{Introduction}

Photonic cluster states, a special class of graph states where many single photons are entangled in a lattice geometry in the graph state representation~\cite{Raussendorf2001}, are key resources for optical quantum computing and quantum communication~\cite{Bartolucci2021, Russo2019, Browne2005}. They pertain to a category called Measurement-Based Quantum protocols, such as the one-way quantum computer or all-optical quantum repeater states~\cite{Raussendorf2001, Azuma2015, Buterakos2017}. Remarkably, a two-dimensional cluster state can be mapped to a conventional quantum computation circuit~\cite{Jozsa2005, Danos2007, Krishnan_Vijayan2024}. 

\noindent Many experimental efforts are currently dedicated to the generation of entangled multi-photon states with the goal of gathering the necessary resources for measurement-based quantum computation (MBQC). The very ambitious goal of generating large two-dimensional states at high fidelity and rate is typically split into the generation of smaller resource states and a few merging operations~\cite{Meng2023Fusion}.  
A first approach to generate small resource states consists in using linear optics \cite{KLM2001} with heralded single-photon sources~\cite{Walther2005, Lu2007, Pilnyak2017}, with a record number of 12 entangled photons obtained on post-selection from 6 sources of photon pairs~\cite{Zhong2018}. Relying on probabilistic photon generation and probabilistic gates, such an approach builds on heavy integration to scale up~\cite{Reimer2019}. 

Conversely, one can deterministically generate linear cluster states from a single quantum emitter and eventually entangle multiple quantum emitters to reach higher dimensionality \cite{Lindner2009}. Such an approach has recently known important progresses with the deterministic generation of linear cluster states with up to $12$ photons with neutral atoms~\cite{PThomas2022, PThomas2024} and up to $4$ particles with semiconductor quantum dots (QDs)~\cite{Schwartz2016, Cogan2021, Coste2023, Su2024, Meng2023, Huet2024}. Generating graph states of higher dimensionality still remains challenging, and optimal architectures will likely combine emitter-based entanglement and linear gates \cite{Hilaire2023neardeterministic}.

Semiconductor quantum dots (QDs) present interesting assets in this regard, having demonstrated both deterministic entanglement generation and unparalleled rates in the generation of photons of high indistinguishability as required for fusion operations~\cite{Somaschi2016, Thomas2021, Tomm2021, Ding2023}. In the present work, we push the generation of linear cluster states with a QD single-photon source and linear gates up to 6 photons and present a method allowing live monitoring of the entanglement scheme. We exploit a resource-efficient linear entanglement scheme based on a compact entangling fiber-based apparatus \cite{Pilnyak2017, Istrati2020}. A fiber-loop delay is used to entangle photons emitted successively by the QD. We produce with the same setup cluster states of $2$, $3$, and $4$ photons, with $4$-photons rates of $2.2$\,Hz and a fidelity lower bound of $63.8\%$. Our high rate allows us to push the entanglement scheme to 6 photons with a stabilizer visibility of $45\pm5\%$, and an entanglement length of 6 photons. Finally, we demonstrate an efficient and fast new method to monitor the entanglement process. This method is based on partially post-selecting $2$, $3$, or $4$-photon events in a measurement where $6$ photons are sent into the apparatus. The high rates of these events allow for live monitoring and optimization of longer cluster state generation all along the measurement. Such a method is an important asset for MBQC protocols where photonic cluster states are constantly generated and measured at a relatively low rate.

\section{Sequential entanglement scheme}\label{scheme}

\begin{figure*}[t]
    \centering
    \includegraphics[width=0.99\linewidth]{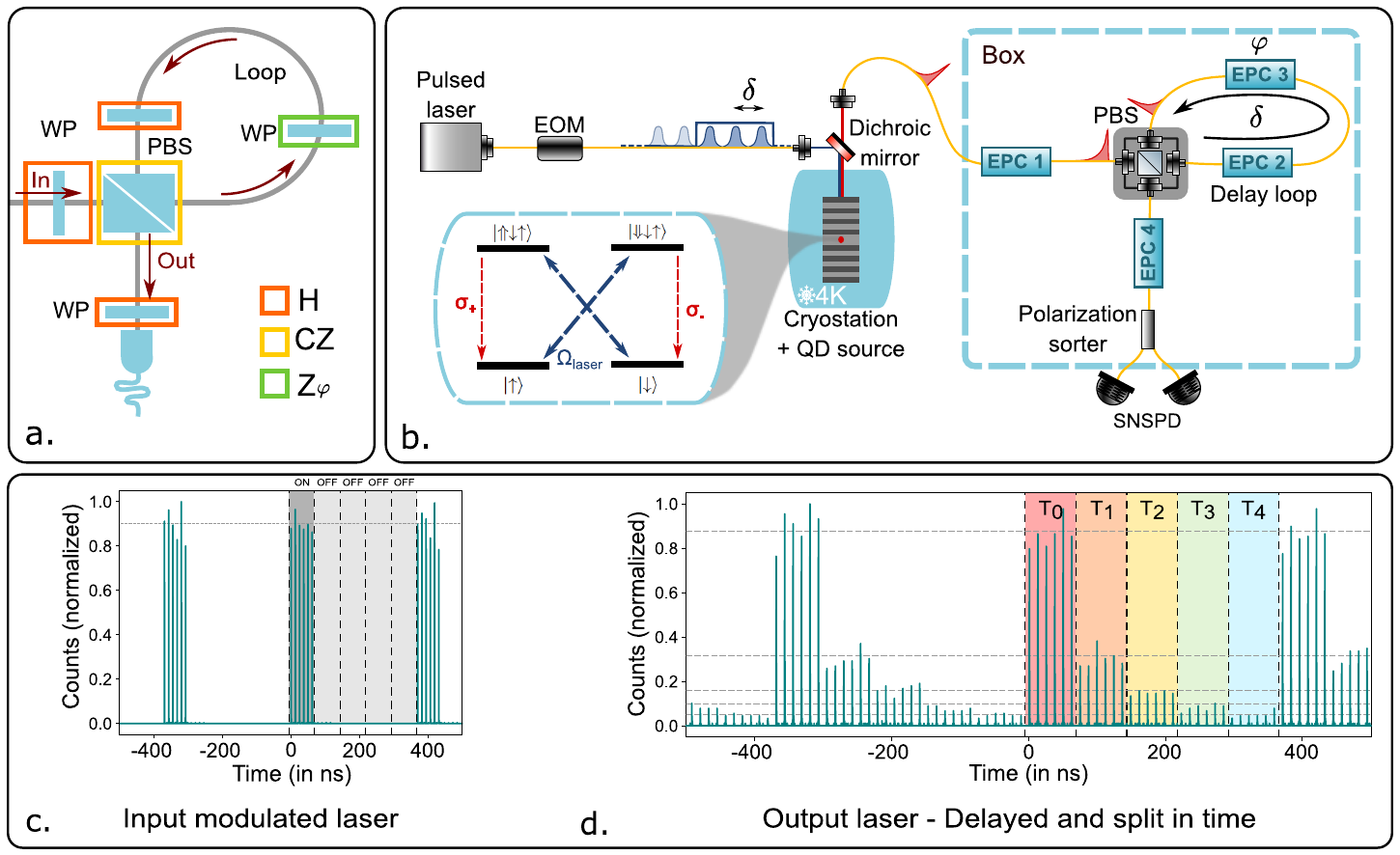}
    \caption{a) The entangling scheme: input single photons are entering sequentially, being stored in the delay loop before being entangled with the next input photon on the PBS by a CZ gate realized by detection post-selection on one of its outputs. b) Experimental setup for the sequential entanglement of single photons from a quantum-dot source. A laser pulse picked by the EOM excites an InGaAs QD in a micropillar cavity, emitting single photons. The photons are entangled sequentially on a PBS connected in a fiber loop configuration. The loop system is mounted in a 19” rack-mountable 2U box, and the polarization of the propagating photons is controlled using four EPCs. The generated cluster state exits the box toward two SNSPDs. Inset: Energy level structure of the negative trion device - Red - Optical transitions of the QD - Blue - Laser drive with $H$ or $V$ polarization. Correlations between the modulated laser and the clock of the experiment, c) before and d) after the loop apparatus. The photons are distributed from the input time slot $T_0$ to several sequential time slots $T_N$ by the Hadamard gate in the loop, leading to exponential decay of the signal in d).}
    \label{fig:Fig1}
\end{figure*}

We follow the entanglement scheme proposed by ~\cite{Pilnyak2017} and sketched in Fig.~\ref{fig:Fig1}a. The core components are a fibered delay-loop acting as a quantum memory and a polarizing beam-splitter (PBS), allowing consecutively generated photons to be entangled in their polarization degree of freedom post-selectively with a $50\%$ success rate. By matching the modes of an input photon and the "memory" photon and post-selecting instances where one photon returns to the memory loop while the other exits, we eliminate the which-path information between them, thereby entangling the photon leaving the system with the one staying in the loop. This technique can be repeatedly used to sequentially entangle chains of photons with the same hardware resources for an arbitrarily large linear cluster state.

In more detail, the qubits are encoded in the polarization degree of freedom, with the eigenstates polarization axes defined by the PBS. Horizontal polarization is designated for transmitted photons and vertical polarization for reflected photons. To entangle two photons using a PBS, both are set in an equal superposition of horizontal and vertical polarizations $\ket{P} = \dfrac{1}{\sqrt{2}}(\ket{H} + \ket{V})$. When the photons are indistinguishable except for polarization, post-selecting the PBS output to allow only one photon through each port isolates two probability amplitudes: one where both photons are transmitted with the $\ket{H}$ polarization to ports 1 and 2 respectively - $\ket{H_1H_2}$, and another where both photons are reflected with the corresponding $\ket{V}$ polarizations -  $\ket{V_1V_2}$, resulting in an entangled Bell state $\Phi^+ = \frac{1}{\sqrt{2}}\left(\ket{H_1H_2}+\ket{V_1V_2}\right)$. Applying a Hadamard rotation to one photon transforms this state into a proper two-photon cluster state, realizing a controlled Z (CZ) operation. We note that this process has a success probability $1/2$, which is limited by the post-selection success rate.

It can be shown that performing this fusion between the last photon of a linear cluster state and an isolated photon in a state $\ket{P}$ attaches the isolated qubit to the end of the linear cluster state, effectively extending it by one qubit~\cite{Pilnyak2017}.
By connecting one of the PBS outputs to one of its inputs, we create the memory loop mentioned above.
When single photons enter the setup sequentially in $N$ time slots in the state $\ket{P}$, while a single photon already exists in the loop, the post-selection condition of one photon exiting each port of the PBS is equivalent to only one photon exiting the loop in each of the $N$ time slots. Each time slot is thus a qubit mode with a polarization degree of freedom.
The experiment starts with inserting a photon into the memory. The photons are entangled sequentially through post-selection. Finally, the memory photon also exits the loop.

We implement the presented scheme using an on-demand single-photon source based on an InGaAs QD deterministically embedded in a micropillar cavity (see supplementary section~\ref{sup:Source} for details on the device structure). The source is in a $4$K cryostat \cite{Somaschi2016, Nowak2014}. The QD confines an electron, with energy levels and optical selection rules displayed in Fig. \ref{fig:Fig1}b. The QD is optically excited using longitudinal acoustic (LA) phonon-assisted excitation \cite{Thomas2021}, with a $15$ps pulsed laser detuned at $\Delta=0.8$nm from the QD resonance and both spectral and polarization filtering is done to get highly polarized pure single photons with a measured $g^{(2)}(0)$ of $2.5\%$. The probability to obtain a single photon at the first lens for each excitation pulse (first lens brightness) is $\eta_\mathrm{FL}=35\%$ and at the output of a single-mode fiber $\eta_\mathrm{SP}=13\%$. The photons also show high Hong-Ou-Mandel visibility $V_\mathrm{HOM}^{12\mathrm{ns}}=87.5\%$ between consecutively emitted photons \cite{Hong1987, Santori2002}, corresponding to a single-photon indistinguishability of $M = 92.3\%$ \cite{Ollivier2020}.

Figure~\ref{fig:Fig1}b also presents the full experimental setup. A Ti:Sapphire pulsed laser with a repetition rate of 81MHz is used to excite the single-photon source around $925$nm.
The laser pulses are modulated by an EOM such that the ON time of the pulse sequence sets the number of photons entangled by the loop, and the OFF time sets a relaxation time for the loop memory before the next sequence (the last photon has a probability $0.5$ to exit the loop after each OFF cycle, decreasing to an overall probability of $0.04$ for $3$ OFF cycles chosen in the following measurements including the losses in the loop).

To overcome the dead time of the detectors, the photons to be entangled are separated by more than one repetition rate. This is done by matching the loop length to 6 laser periods, which makes the entangled photon sequences interleaved - the first photon is entangled to the 7th, then to the 13th, and so on, so that every 6 consecutive photons belong to 6 different trials of the experiment.
The entangling part of the setup is composed of a fiber-coupled PBS and four Electric Polarization Controllers (EPC) that are used to control the polarization of the photons in the fiber.
The polarization of the incoming photons is initialized by EPC1 to $\ket{P}$. The PBS acts as a CZ gate between the memory and input photons (see Fig.~\ref{fig:Fig1}a). In the loop, two EPCs control the polarization of the photons to implement two functionalities that are required inside the loop -- a Hadamard rotation, which is part of the sequential entangling sequence, and an additional birefringent phase $Z_{\varphi}$ enabling a visibility scan for state characterization.
Figure~\ref{fig:Fig1}c  presents an input sequence with one time slot ON and four time slots OFF. Each ON time slot is comprised of 6 photon pulses belonging to separate experimental trials. Fig.~\ref{fig:Fig1}d  presents the output detection sequence of such an input in one of the detectors. The input photons are first split by the PBS (due to the input Hadamard), half are reflected and measured without delay in time slot $T_0$ without entering the loop, while the rest are delayed by the loop and split by the PBS once more in each round with reduced amplitude caused by the Hadamard rotation inside the loop and the accumulated losses in each round~\cite{Steindl2021}. This leads to exponential relaxation of the memory loop, evidenced by the signal in Fig.~\ref{fig:Fig1}d measured in time slots $T_1$ to $T_4$.
A 2-photon experiment would show 12 consecutive pulses in the input plot, and an intricate pattern on the output.

Finally, at the output of the fibered setup, EPC4 sets the polarization detection basis to $X$ and the polarization is projected to two nano-wire detectors by a fused in-fiber polarization sorter.
The phase operation $Z_{\varphi}$, although applied inside the loop, allows to rotate non-locally the measurement basis outside the loop from $X$ to $X_{\varphi}=Z_{\varphi}XZ^{\dagger}_{\varphi} = \begin{pmatrix}
  0 & e^{-i\varphi}\\ 
  e^{i\varphi} & 0
\end{pmatrix}$. This observable resides on the $X-Y$ equator of the Bloch sphere and allows us to measure, for example, both X and Y. 
The first $N-1$ photons that exit the loop are projected and measured in this way. Conversely, the polarization of the last photon that is in the loop is projected by the loop PBS as the PBS no longer serves the purpose of an entangling operation. Thus, the last photon is always measured in the $Z$ basis. 
Overall, this setup can measure observables of the form
\begin{equation}\label{eq.MeasurementObservables}
    O(\varphi) = X_{\varphi}^{1}X_{\varphi}^{2}..X_{\varphi}^{N-1}Z^{N}\,.
\end{equation}
Here $X^n$ is the $X$ Pauli operator on the $n_{th}$ photon, and each of the single-photon Pauli operators can also be replaced by the unit operator $I$.

To characterize the state and derive lower bounds on the entanglement level of the produced state, measurements are conducted by scanning $\varphi$ and measuring the mean value of different observables. 

\begin{equation}\label{eq.visibility}
    V(\varphi) = \mathrm{Tr}(\rho\cdot O(\varphi))=\braket{O(\varphi)}
\end{equation}

For certain values of $\varphi$, an observable may be a stabilizer of the state, for which the measured value would reach a maximal value of $1$ in an ideal case.

\section{Entanglement up to 6 photons}

\begin{figure}[h!]
    \centering
    \includegraphics[width=0.95\columnwidth]
    {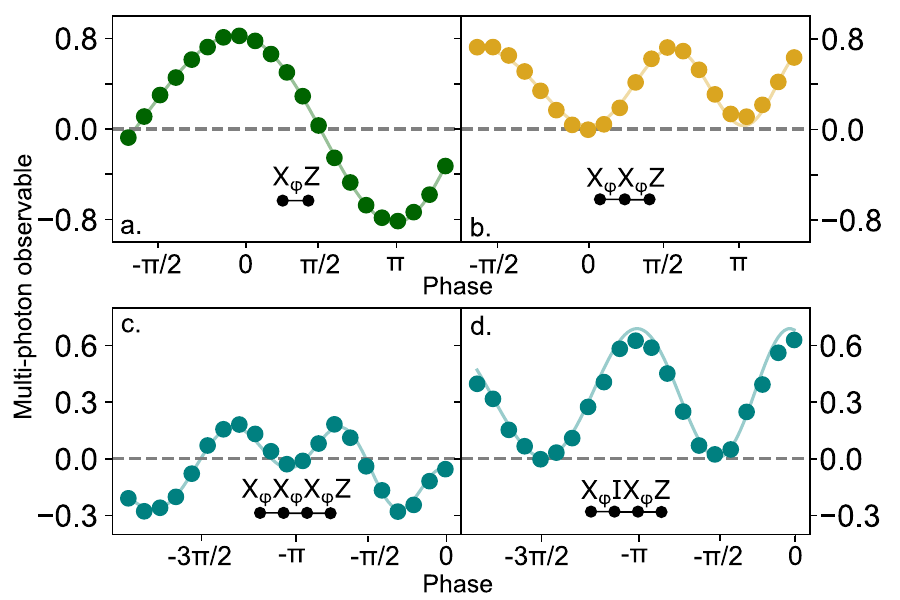}
\caption{Measured $N=2$ to $4$-photon observables, characterizing $N-$photon cluster state generation. The experimental data correspond to the symbols with smaller error bars than symbol size, and the line represents the expected behavior derived from $V_\mathrm{HOM} = 82.7\pm1.6\%$. The measured observable is in the inset for a) a $2$-photon cluster state, b) $3$-photon cluster state, c) $4$-photon cluster state for the visibility $V_4$, and d)  $V_4'$.}
\label{fig:Visibilities}
\end{figure}

The entanglement scheme was used to experimentally entangle 2, 3, 4, and 6 single photons. For each $N$-photon scenario, we control the EPC3 phase $\varphi$ and measure the accessible observable. Fig.~\ref{fig:Visibilities} presents such measurements obtained for 2, 3 and 4 photons for the observables $O_2(\varphi) = X_{\varphi}Z$, $O_3(\varphi) = X_{\varphi}X_{\varphi}Z$, $O_4(\varphi) = X_{\varphi}X_{\varphi}X_{\varphi}Z$, and $O_4'(\varphi) = X_{\varphi}IX_{\varphi}Z$.  Each curve shows oscillatory behavior as a function of $\varphi$, for which we can extract visibilities $V_2$, $V_3$, $V_4$, and $V_4'$ corresponding to the normalized difference between the highest measured value to the lowest (oscillations from 0 to 1) or mean (oscillations from -1 to 1) value.

First, we focus on the quality of entanglement of two photons quantified by the $\braket{X_{\varphi}Z}$ observable. This observable oscillates between the stabilizer $\braket{XZ}$ and the $\braket{YZ}$ observable, of respective eigenvalues $+1$ and $-1$ for an ideal $2$-photon cluster state. Since the 2-photon state is prepared through a single entanglement operation, the visibility is theoretically expected to follow  $V_2(\varphi)=V_2 \cos(\varphi)$ with $V_2=V_\mathrm{HOM}$~\cite{Istrati2020}. The symbols in Fig.~\ref{fig:Visibilities}a present the measured visibility and the line presents the expected visibility considering the independently measured HOM visibility $V_\mathrm{HOM} = 82.7\pm1.6\%$, which probes the interference level between two photons separated by $73.9$\,ns (see supplementary section \ref{sup:HOM} for details on the measurement), showing excellent agreement with experimental observations.

Now we entangle one more photon and measure $\braket{X_{\varphi}X_{\varphi}Z}$, which oscillates between $\braket{YYZ}$ and $\braket{XXZ}$ with respective eigenvalues of $1$ and $0$ for an ideal $3$-photon cluster state. Since this measure is the result of a stabilizer measurement in $X$ basis for two photons, the visibility is expected to follow $V_3(\varphi)=V_3\sin^2(\varphi)$ with $V_3=V_\mathrm{HOM}^2$~\cite{Istrati2020}. The symbols in Fig.~\ref{fig:Visibilities}b  follow the expected trend: it is shifted by $\pi/2$ with respect to $\braket{X_{\varphi}Z}$ and oscillates twice as fast. Here again, the comparison with the theoretically expected behavior shows excellent agreement with measurements.

We now turn to 4-photon entanglement and measure two observables: $\braket{X_{\varphi}X_{\varphi}X_{\varphi}Z} =  V_4(\varphi)= -V_4 \cos(\varphi) \sin^2(\varphi)$ with $V_4=V_\mathrm{HOM}^3$ which for an ideal $4$-photon cluster state oscillates between $\pm\frac{2}{3\sqrt{3}}$ as well as $V_4^{'}(\varphi)=\braket{X_{\varphi}IX_{\varphi}Z} = V_4^{'} \cos^2(\varphi)$ with $V_4^{'}= V_\mathrm{HOM}^2$ which oscillates between $0$ and $1$. Experimental data in Figs. ~\ref{fig:Visibilities}c and d show excellent agreement with the expected values. Only a slight reduction of contrast compared to the expectations is observed for $V_4^{'}$. Several experimental reasons can explain such reduction, such as small variations of the photon indistinguishability due to residual electric noise, imperfect extinction of the laser pulses by the EOMs, misalignment, and imperfect extinction ratio of the PBS, as well as incomplete emptying of the memory loop. We fitted the experimental curves to extract the visibilities up to 4 photons, finding $V_2 = 82.0\pm0.2\%$, $V_3=74.4\pm1.0\%$, $V_4=59.7\pm1.2\%$ and $V_4^{'} = 62.2\pm1.0\%$. These values allow us to deduce a lower bound on the fidelities for the three generated cluster states~\cite{Istrati2020} of $F_2 \geq 86.5\pm0.1\%$, $F_3 \geq 77.6\pm0.9\%$, and $F_4 \geq 63.8\pm1.0\%$ for two, three, and four photons, respectively. These fidelities are above the $50\%$ limit by several standard deviations, demonstrating genuine entanglement.

When comparing our results with the previous state-of-the-art values using the same linear optics-based entangling method~\cite{Istrati2020}, we observe both improved visibilities and detection rates. Concretely, the visibilities are higher by $9\%$, $5\%$, and $4\%$ for $N = 2, 3\text{ and }4$ photons, respectively, with detection rates of $6\,k\mathrm{Hz}$, $120\,\mathrm{Hz}$ and $2.2\,\mathrm{Hz}$, corresponding to an improvement factor of $12$, $28$, and $55$ of the detection rates. This is due to the higher indistinguishability of our single photons and to the reduced optical losses in the experiment. These losses can be analyzed into a scaling ratio, defined in Eq.~\ref{eq:scaling} \cite{Istrati2020}, which measures the reduction in the detection rate of an $N$-photon cluster state as an additional photon is added to the cluster. This metric encompasses all system efficiencies and is defined as
\begin{equation}
    r=\frac{R_N}{R_{N+1}}=\dfrac{1}{\eta_\mathrm{SP} \eta_\mathrm{setup} \eta_\mathrm{d} \eta_\mathrm{ent}},
    \label{eq:scaling}
\end{equation}
where $R_N$ is the detection rate of an $N$-photon cluster state, $\eta_\mathrm{SP}$ is the single-photon fibered brightness, $\eta_\mathrm{setup}$ is the loop setup transmission efficiency, $\eta_\mathrm{d}$ is the detector efficiency, and $\eta_\mathrm{ent}$ is the entangling success probability. 
In the current work, $\eta_\mathrm{SP}=13\%$, $\eta_\mathrm{setup}=37\%$, $\eta_\mathrm{d}=90\%$, and $\eta_\mathrm{ent}=50\%$, limited by the probabilistic nature of the entangling gate given by post-selection. This results in a scaling ratio of $46\pm 5$, twice lower than our previously reported value \cite{Istrati2020}.

\begin{figure}[h!]
    \centering
    \includegraphics[width=0.95\columnwidth]
    {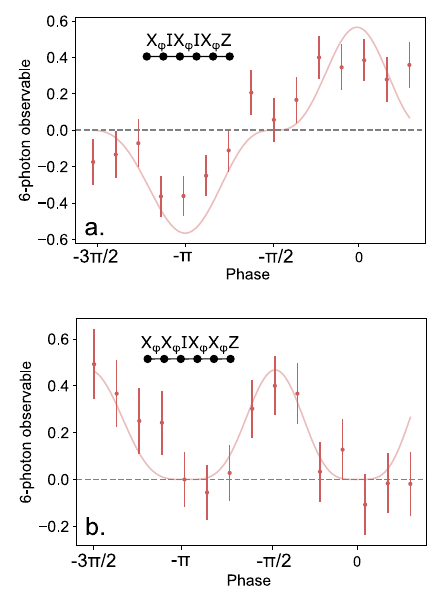}
\caption{Measured $6$-photon observables with the experimental data corresponding to the symbols and the line representing the expected behavior derived from $V_{\mathrm{HOM}} = 82.7\pm1.6\%$. The observables are a) $X_{\varphi}IX_{\varphi}IX_{\varphi}Z$ corresponding to $V_6^\prime$ and b) $X_{\varphi}X_{\varphi}IX_{\varphi}X_{\varphi}Z$ corresponding to $V_6^{\prime\prime}$.}
\label{fig:6Vis}
\end{figure}

From the value of the two-photon indistinguishability $V_\mathrm{HOM}=82.7\pm1.6\%$, we deduce that our protocol ensures the generation of $N$-photon genuine entanglement in a linear cluster state for $N> 6$~\cite{Istrati2020}. Taking advantage of the improved scaling ratio $r = 46\pm 5$, we thus pushed the entanglement scheme to $6$-photon measured at a rate of $2.2$\,mHz. Here, we measured the visibilities corresponding to three observables. The observable $\braket{X_{\varphi}X_{\varphi}X_{\varphi}X_{\varphi}X_{\varphi}Z}$  corresponds to a theoretical visibility 
$V_6=-V_\mathrm{HOM}^5 \cos^3(\varphi) \sin^2(\varphi)$,  varying between $\approx \pm 18.5 \%$ for unity photon indistinguishability and $\approx\pm 8\%$ for the current $V_\mathrm{HOM}$. The corresponding experimental measurement (shown in supplementary Fig.~\ref{fig:6-Vis-Solo} together with the theoretical expectation) shows that our signal-to-noise ratio does not allow to resolve the expected oscillations.  However, we report in Fig.~\ref{fig:6Vis} two visibilities measured within the same experimental run. In Fig.~\ref{fig:6Vis}a, $V_6^{'}$ corresponds to the observable $\braket{X_{\varphi}IX_{\varphi}IX_{\varphi}Z}$ oscillating between $\braket{XIXIXZ}$ and $\braket{YIYIYZ}$ with eigenvalues of $1$ and $-1$. In Fig.~\ref{fig:6Vis}b,   we present the  visibility $V_6^{''}$ of the observable  $\braket{X_{\varphi}X_{\varphi}IX_{\varphi}X_{\varphi}Z}$, oscillating between $\braket{YYIYYZ}$ and $\braket{XXIXXZ}$ with eigenvalues of $1$ and $0$. The corresponding theoretical curves given by $V_6^{'}(\varphi)=V_\mathrm{HOM}^4 \cos^{3}(\varphi)$ and $V_6^{''}(\varphi)=V_\mathrm{HOM}^4 \sin^{4}(\varphi)$ evidence an excellent agreement between observations and theory, further supporting the presence of genuine entanglement for 6 photons.

\section{Partial post-selection measurement}
For a practical application of MBQC, any protocol must reliably generate large cluster states over extended periods. Monitoring the system stability and alignment over long durations without requiring maintenance that could disrupt the computation is essential. Additionally, the ability to discard data compromised by technical issues without interrupting the experiment is crucial. To address these aspects, we propose a partial post-selection visibility measurement, which can be obtained at a relatively higher rate than the main signal.

In a $N$-photon experiment, $N$ photons are prepared at the input of the experiment, propagate through the setup, and exit toward the detectors. The post-selection relies on selecting only events with $N$ photons detected in a specific timing arrangement. This typically requires a long acquisition time to accumulate enough statistics. During long acquisition times, performance indicators are required to ensure the proper functioning of the protocol. Due to loss, the output of the setup exhibits $M$-photon events, with $M < N$, with higher rates. The \textit{Partial Post-Selection} (PPS) method uses these inadequate events to obtain information about the experiment. For example, if the last input photon is lost, the experiment becomes effectively a $(N-1)$-photon experiment, which should, in principle, produce $N-1$ long cluster states. However, other photons can be lost, leading to a reduced cluster state length, hence also $N$-cluster state visibility. As an example, the various contributions of these partial post-selected events are shown for $M=2$ and $N=3$ in the supplementary section ~\ref{sup:Pattern}.

\begin{figure}[h!]
    \centering
    \includegraphics[width=\columnwidth]{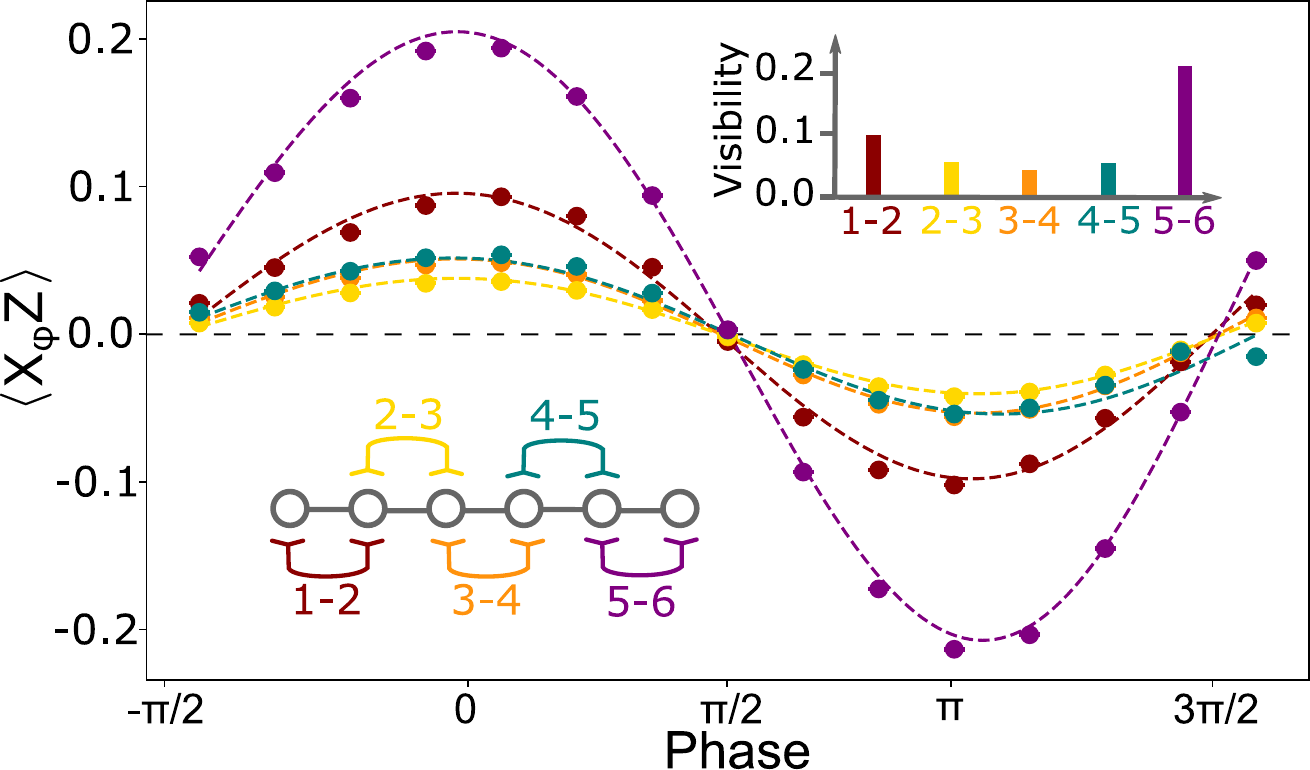}
    \caption{
    Partially post-selected 2-photon observable measurement plots in a 6-photon experiment. The color corresponds to the position of the 2 photons measured within the 6-photon chain. The error bars are plotted but smaller than the marker size. The oscillations are fitted with the same function as for the 2-photon measurement ${V}_{2,\mathrm{PPS}}(\varphi) = V_{2,\mathrm{PPS}}\cos(\varphi+ \varphi_0)$. Inset: Visibility $V_{2,\mathrm{PPS}}$ of the oscillations plotted as a function of the position in the cluster state.}
    \label{fig:SlidePPS}
\end{figure}

Using time analysis, it is possible to post-select the $M$ photons at different positions in the $N$-photon cluster state. By changing the post-selection, we effectively vary the source of noise, resulting in varied visibility. Qualitatively, the noisy events can be caused either by photons created prior to the desired $M$-photon cluster state that remained in the loop, or by photons created after which did not enter the loop and were reflected to the detectors. Fig.~\ref{fig:SlidePPS} illustrates the measurement of the 2-photon observable (with a $5$\,kHz rate) along the 6-photon linear cluster state (at $2$\,mHz rate). The color indicates the position within the cluster (as depicted in the inset), with the visibility of each curve varying according to the position, with a maximum at $V_{2,\mathrm{PPS}} = 22.00\pm0.08\%$. The corresponding visibilities are displayed in the top right inset. As presented in Fig.~\ref{fig:SlidePPS}, post-selecting the first and last photons of the cluster results in improved visibility due to the reduced noise from early and late photon events. Post-selecting in the middle of the chain has a similar noise anywhere, thus, similar visibility. In the two best cases, measuring the last photons of the cluster state, instead of the first, improves the visibility measured. This is due to the noise from late photons reflected by the loop PBS, which has a larger probability amplitude than the noise from early photons staying in the loop. This observation is true for any $M$-photon PPS, thus, we always use the last photons of the cluster to perform PPS measurements.

\begin{figure}[t]
    \centering
    \includegraphics[width=0.45\textwidth]{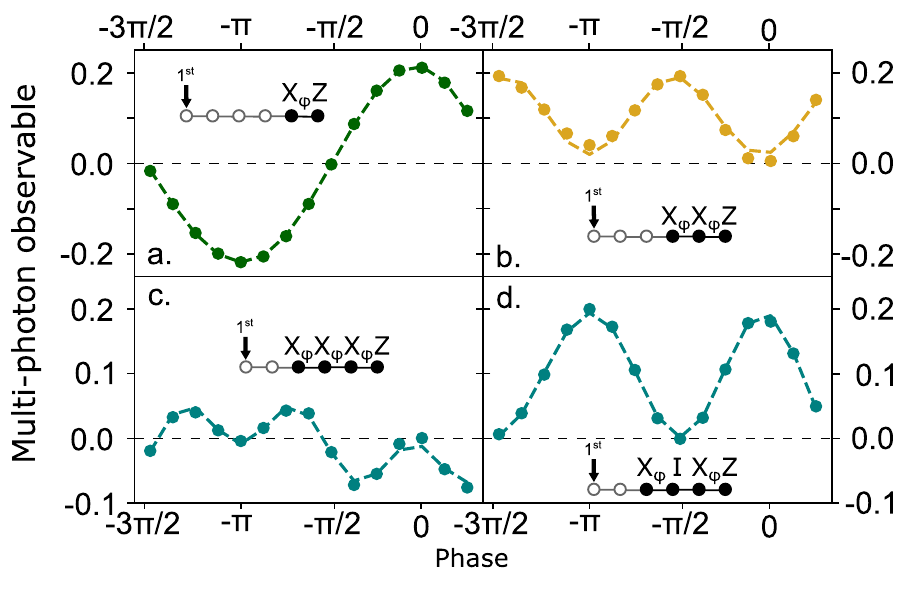}
    \caption{Partially post-selected measurements of the observables a) $X_{\varphi}Z$, b) $X_{\varphi}X_{\varphi}Z$, c) $X_{\varphi}X_{\varphi}X_{\varphi}Z$, and d) $X_{\varphi}IX_{\varphi}Z$ in a 6-photon experiment. They follow the identical oscillatory behavior as $V_{2,\mathrm{PPS}}(\varphi)$, $V_{3,\mathrm{PPS}}(\varphi)$, $V_{4,\mathrm{PPS}}(\varphi)$, $V_{4,\mathrm{PPS}}^{'}(\varphi)$ with lower amplitudes than the observables shown in Fig.\ref{fig:Visibilities}. The dotted lines represent the fits. Inset: The black points represent the position of the photons measured in the $6$-photon cluster state.}
    \label{fig:Residuals}
\end{figure}

Figure \ref{fig:Residuals} shows PPS measurements from the $6$-photon experiment.  Post-selection on $2$, $3$ and $4$-photon events yielded the corresponding plots. These curves are similar to those of the corresponding measurements plotted in Fig.~\ref{fig:Visibilities}, with phase dependence corresponding to the $2$, $3$, and $4$-photon interferences as expected from the measured observables. The reduced amplitudes are due to the polluting output patterns being post-selected along with the ones generating the cluster states. The more photons in the PPS visibility measurement, the more combinations of polluting outputs reduce the PPS visibility, e.g. for 4 photons more than for 2 photons. The amplitudes of the $2$, $3$, and $4$-photon PPS visibilities are $V_{2,\mathrm{PPS}}=22\%$, $V_{3,\mathrm{PPS}}=19\%$, and $V'_{4,\mathrm{PPS}}=19\%$, respectively, enough to monitor the value with errors smaller than the marker size. The acquisition times per point are $10$ seconds, $90$ seconds, and $15$ minutes, respectively.

Such PPS allows to continuously maximize a $M < N$ visibility measurement during a long experiment, to ensure the best performance without interrupting the measurement. For example, to measure $6$-photon interferences, we repeatably scanned the phase instead of measuring much longer at a pre-set phase. During the scans, we monitored detection count rates and $M$-photon visibilities, enabling us to correct for experimental drifts. Explicitly, to concatenate all the acquired scans together, we used partially post-selected visibilities to correct for small phase drifts in the loop fiber between individual phase scans caused by EPC3 small hysteresis, effectively changing the initial phase $\varphi_0$ of the fit. Such procedure allows us to extract the relative phase information determined from the fitting of the individual 2-photon PPS measurements with $V_{2,\mathrm{PPS}}\cos(\varphi+\varphi_0)$, shown in Fig.~\ref{fig:Residuals}a. This led to a more accurate calculation of the mean and standard deviation of the phase value at each point of the 6-photon scan. Due to the fast acquisition of the PPS, 2-photon visibilities, inherently present in every $N$-photon experiment, provide the ideal tool for live monitoring of the generation of long cluster states.

\section{Conclusion}

In conclusion, we demonstrated linear cluster state generation using sequentially emitted single photons from a single quantum-dot device and their entanglement using a linear optical setup. We use this entanglement scheme with a solid-state single-photon source and, for the first time, entangle up to six photons with mHz measured rate, resulting from an improved setup efficiency. The results can be further improved by reducing the losses even more and using the best single-photon source reported to date \cite{Ding2023} with up to $\eta_\mathrm{SP}=71\%$ end-to-end efficiency and $V_\mathrm{HOM}=94\%$. 
Although limited by the fundamental probabilistic nature of the entangler, this method is highly relevant for building increasingly larger cluster states from smaller resource states~\cite{Bartolucci2021} for MBQC applications.
Additionally, we introduced the concept of PPS visibility measurement, which leverages low-photon-number states in large cluster state generation experiments to monitor and perform these experiments while ensuring the quality of the generated states and the stability of the experiment. We anticipate that this method is not specific to our experimental protocol and may apply to other implementations using linear optical entanglement \cite{Pont2022} as well as other deterministic protocols \cite{PThomas2022}.

\section*{Acknowledgements}
This work was partially supported by the Paris Ile-de-France Région in the framework of DIM QUANTIP, the French Defense ministry - Agence de l'innovatio de défense, the European Union’s Horizon CL4 program under the grant agreement 101135288 for EPIQUE project, the Plan France 2030 through the projects ANR22-PETQ-0011, ANR-22-PETQ-0006, and ANR-22-PETQ-0013, and a public grant overseen by the French National Research Agency as part of the ”Investissements d’Avenir” programme (Labex NanoSaclay, reference: ANR10LABX0035). This work was carried out within the C2N micro-nanotechnologies platforms and was partially supported by the RENATECH network and the General Council of Essonne.

\section*{Data and materials availability} All data that support the findings of this study are included within the article (and any supplementary files).

\section*{Competing interest} P.S. (Pascale Senellart) is a scientific advisor and co-founder of the company Quandela. The other authors declare no competing interests.

\bibliographystyle{unsrtnat}
\bibliography{Bibliography}

\clearpage

\setcounter{section}{0}
	\setcounter{equation}{0}
	\setcounter{figure}{0}
	\setcounter{table}{0}
	\setcounter{page}{1}
	\makeatletter
	\renewcommand{\thetable}{S\arabic{table}}
	\renewcommand{\thesection}{S-\Roman{section}}
	\renewcommand{\theequation}{S\arabic{equation}}
	\renewcommand{\thefigure}{S\arabic{figure}}

\begin{center}
	\textbf{\large SUPPLEMENTARY INFORMATION AND DATA}
\end{center}

\section{Single-photon device}\label{sup:Source}

Our single-photon source is based on an InGaAs quantum dot (QD) deterministically embedded in a micropillar cavity \cite{Somaschi2016, Nowak2014}. The cavity is formed by two distributed Bragg mirrors (DBR): the top (bottom) DBR is composed of $16$ ($34$) pairs of alternating AlAs/GaAs layers. The top (bottom) DBR is p (n) doped and electrically contacted, forming a p-i-n junction. The QD is grown in the junction intrinsic region, enabling us to apply reverse bias to tune the QD in the cavity resonance and to trap a negative charge in the QD.

\section{Loop-delay Hong-Ou-Mandel measurement} \label{sup:HOM}

\begin{figure}[h!]
    \centering
    \includegraphics[width=\columnwidth]{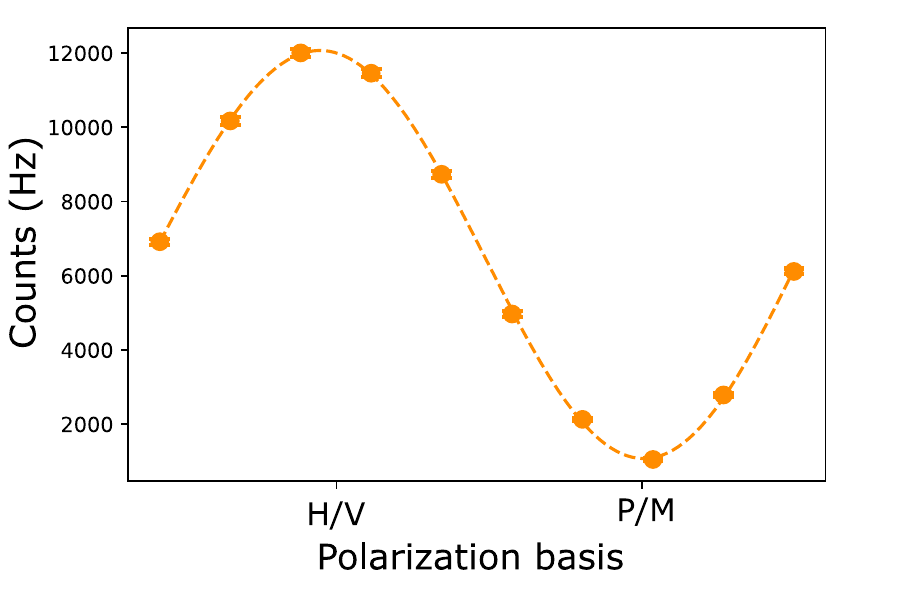}
    \caption{
    Hong-Ou-Mandel visibility measurement in the delay-loop apparatus with $V_{\mathrm{HOM}}^{} = 82.7\%$. Experimental data points are fitted with $V_{\mathrm{HOM}}^{}\cos(x)$ (dotted line).}
    \label{fig:HOM}
\end{figure}

To quantify $V_{\mathrm{HOM}}^{ }$ for the precise entangling loop delay, we programmed the EOMs and conducted a two-photon pulse sequence experiment. The single photons are prepared with EPC1 in $\ket{P}$ polarization. The first photon is, with equal probability, either reflected or transmitted. If the first photon is reflected, this situation is discarded by the timing analysis and does not contribute to two-photon coincidences used to determine $V_{\mathrm{HOM}}^{ }$. When the first photon is transmitted into the loop, it arrives at the entangling PBS simultaneously with the second photon. Since they both arrive with $\ket{P}$ polarizations, the PBS routes them according to their projection in $\{H, V\}$ basis. One of the four equally probable outcomes is that both photons leave toward the polarization sorter in the detection when one photon has horizontal and one has vertical polarization. This case of two photons exiting the PBS together is disregarded in the main entangling scheme, but it is very useful for this indistinguishability measurement. Therefore, by adjusting the polarization basis of the tomography setup and scanning the polarization of EPC4, we can alter the intensity of the photon bunching, as shown in Fig.~\ref{fig:HOM}. Concretely, for tomography basis aligned to $P/M$ polarization, we erase the \textit{which path} information and observe the maximal bunching, corresponding to $V_{\mathrm{HOM}}^{ }=82.7\%$. In contrast, when using the $H/V$ basis, the photons become distinguishable by their polarization, which reveals whether or not they entered the delay loop. 

\section{6-photon visibility} \label{sup:6-Vis}

\begin{figure}[h!]
    \centering
    \includegraphics[width=0.95\columnwidth]
    {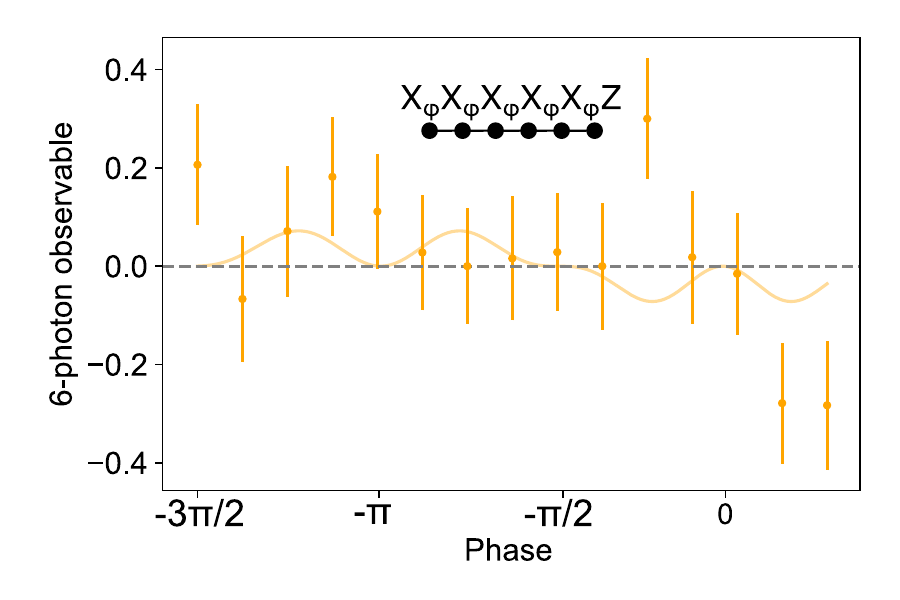}
\caption{$6$-photon measurements of the observable $X_{\varphi}X_{\varphi}X_{\varphi}X_{\varphi}X_{\varphi}Z$ follows $V_6(\varphi) = -V_6\cos^{3}(\varphi)\sin^{2}(\varphi)$, $V_6=V_\mathrm{HOM}^5$. The line represents the expected oscillation from $V_\mathrm{HOM}^\mathrm{ } = 82.7\pm1.6\%$.}
\label{fig:6-Vis-Solo}
\end{figure}

Fig.~\ref{fig:6-Vis-Solo} presents the measurement of the $X_{\varphi}X_{\varphi}X_{\varphi}X_{\varphi}X_{\varphi}Z$ in the same experimental run as the data reported in Fig.~\ref{fig:6Vis}. The line presents the expected behavior $V_6(\varphi) =-V_\mathrm{HOM}^5 \cos^3(\varphi) \sin^2(\varphi) $, evidencing oscillations expected to vary within $\approx\pm 8\%$ with the current $V_{\mathrm{HOM}}$ ($\pm18.6\%$ in the perfect case). The experimental data do not allow to evidence such a visibility.  Strong support for entanglement is given in the main text with measurements of $X_{\varphi}IX_{\varphi}IX_{\varphi}Z$ and $X_{\varphi}X_{\varphi}IX_{\varphi}X_{\varphi}Z$.

\section{Post-selection pattern} \label{sup:Pattern}

\begin{figure}[h!]
    \centering
    \includegraphics[width=\columnwidth]{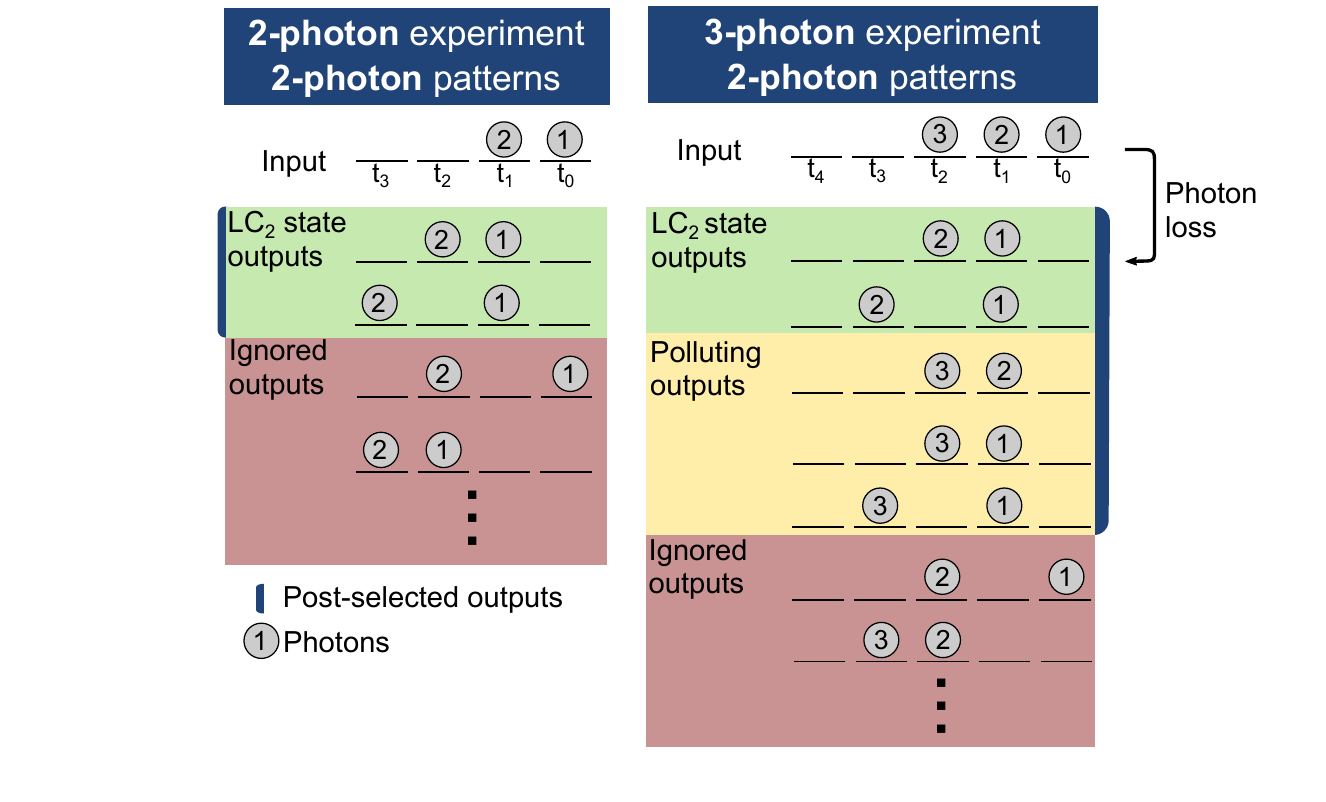}
    \caption{
    Right: Post-selection pattern for a $2$-photon experiment. Left: Partial Post-Selection (PPS) pattern for a $2$-photon observable measurement in a $3$-photon experiment.
    }
    \label{fig:PPSPattern}
\end{figure}

We illustrate here the protocol of partial post-selection for $M=2$ photons in a $N=3$ photon experiment. In Fig.~\ref{fig:PPSPattern} (left), we show the post-selection pattern realized on a $2$-photon experiment. With two photons in the input of the experimental setup, several output configurations are possible. The numbering we use is not connected to the emission order but to the photon ordering at the entangling setup exit. Explicitly, the first photon leaving the entangling setup can be the photon emitted first, which spent time in the delay loop, or the second one, which did not enter the loop. We can not distinguish these photons because they are entangled. To satisfy the condition of entanglement, both photons need to be delayed by the loop once or twice for the second  (green box) in the example in Fig.~\ref{fig:PPSPattern}. Any other situation does not lead to entanglement and is therefore discarded within the post-selection (red box).

We apply the $2$-photon experiment post-selection rules to the $3$-photon experiment. In this experiment, we retrieve information even if one photon is lost. For example, if the last photon of the input is lost, the situation reduces exactly to the $2$-photon experiment. However, in Fig.~\ref{fig:PPSPattern} (right), we can see that there are other $2$-photon events possible (yellow box), not corresponding to any cluster state, which reduces the visibility of the $2$-photon measurement.

\end{document}